\documentclass[sigconf]{acmart}

\setcopyright{none} %
\renewcommand\footnotetextcopyrightpermission[1]{}

\usepackage{multirow,tabularx}
\usepackage{upquote}
\usepackage{xspace}
\usepackage{comment}
\usepackage{tikz}
\usepackage{graphicx}
\usepackage{subcaption}
\usepackage{caption}
\usepackage{enumitem,kantlipsum}
\usepackage{tcolorbox}

\newcommand{\sysname}{{{\fontfamily{lmss}\selectfont GrAALF}}\xspace}

\begin{document}

\title{
\sysname :Supporting Graphical Analysis of Audit Logs for Forensics
}

\author{Omid Setayeshfar}
\affiliation{%
  \institution{University of Georgia}
  \city{Athens}
  \state{Georgia}
  \postcode{30602}
  }
\email{omid.s@uga.edu}

\author{Christian Adkins}
\affiliation{%
  \institution{University of Georgia}
  \city{Athens}
  \state{Georgia}
  \postcode{30602}
}
\email{caadkins@uga.edu}

\author{Matthew Jones}
\affiliation{%
  \institution{University of Georgia}
  \city{Athens}
  \state{Georgia}
  \postcode{30602}
}
\email{matthew.jones@uga.edu}

\author{Kyu Hyung Lee}
\affiliation{%
  \institution{University of Georgia}
  \city{Athens}
  \state{Georgia}
  \postcode{30602}
}
\email{kyuhlee@cs.uga.edu}

\author{Prashant Doshi }
\affiliation{%
  \institution{University of Georgia}
  \city{Athens}
  \state{Georgia}
  \postcode{30602}
}
\email{pdoshi@cs.uga.edu}

\renewcommand{\shortauthors}{Setayeshfar, et al.}


\begin{abstract}

System-level audit logs often play a critical role in computer forensics. They capture low-level interactions between programs and users in much detail, making them a rich source of insight and provenance on malicious user activity. However, using these logs to discover and understand malicious activities when a typical computer generates more than 2.5 million system events hourly is both compute and time-intensive.

We introduce a graphical system called \sysname for efficiently loading, storing, processing, querying, and displaying system events to support computer forensics. In comparison to other related systems such as AIQL~\cite{gao2018aiql} and SAQL~\cite{gao2018saql}, \sysname offers the flexibility of multiple backend storage solutions, easy-to-use and intuitive querying of logs, and the ability to trace back longer sequences of system events in (near) real-time to help identify and isolate attacks. Equally important, both AIQL and SAQL are not available for public use, whereas \sysname is open-source.

\sysname offers the choice of compactly storing the logs in main memory, in a relational database system, in a hybrid main memory-database system, and a graph-based database. We compare the responsiveness of each of these options, using multiple huge system-call log files. Next, in multiple real-world attack scenarios, we demonstrate the efficacy and usefulness of \sysname in identifying the attack and discovering its provenance. Consequently, \sysname offers a robust solution for analysis of audit logs to support computer forensics.

\end{abstract}

\maketitle

\begin{abstract}

\end{abstract}

\section{Introduction}

Provenance data contained in system logs offers a rich source of information for computer forensics. This is because the log comprehensively captures how processes controlled by an attacker interact with resources such as disk and network. System events loggers such as Linux Audit~\cite{audit}, Sysdig~\cite{sysdig}, DTrace~\cite{dtrace}, and Event Trace for Windows (ETW)~\cite{etw} are often used to generate these logs.
Although logs are frequently analyzed offline after an attack has happened, real-time system monitoring can help the user in various ways: if forensic logs can be analyzed in real-time, this rich source of information allows investigation of ongoing abnormal behaviors and thus can protect the system more effectively. Also, timely attack investigations are essential to protecting the system from similar future attacks.

However, we see three main challenges in developing a practical system that can support (near) real-time analysis.
First, system logs grow rapidly. We observe that a single host generates more than 2.5 million system events in an hour. Other studies~\cite{lee2013loggc,king2003backtrackingIntrusions,goel2005taserIntrusionRecovery} also reported that a single device generates over 3 GB of audit log daily.
This challenge has motivated previous research into both lossless and lossy compression of  logs~\cite{lee2013loggc,ma2016protracer,xu2016highFidelity,hossain2018dependence,tang2018nodemerge} as well as optimizing querying and pattern matching in these logs~\cite{gao2018aiql,gao2018saql}.

Next, there exist various logging systems for different operating systems and underlying architectures, and their definitions of the event entries, as well as output formats, are often different.
For instance, Linux and Unix systems frequently use Linux Audit~\cite{audit} to log system events. On the other hand, Windows systems mostly use Event Tracing for Windows (ETW)~\cite{etw} to record system events, including system calls and Windows API calls.
Linux and Windows have completely different system calls and system APIs. Furthermore, recent studies~\cite{lee2013beep,pohly2012hifi,bates2015trustworthy,muniswamy2006provenance} propose novel logging techniques to improve the effectiveness and efficiency of computer forensics.
For instance, BEEP~\cite{lee2013beep} proposes a technique to divide the long-running process into a finer-grain system object called {\it execution unit} to improve forensic accuracy.  All of this makes it challenging to seamlessly support forensics on system logs generated by various logging platforms.

Third, backward and forward tracking techniques~\cite{king2003backtrackingIntrusions, goel2005taserIntrusionRecovery} are essential in computer forensics to understand causal relations between system objects (e.g., process) and subjects (e.g., file, network socket). They often require tracking back to previous events to identify causal chains. In practice, database solutions are often used to store sequences of system events, and the user composes queries to identify causal relations between system components. However, our evaluation results show that existing database solutions cannot provide enough performance to support backward and forward tracking for streaming system events in real-time. 
Specifically, we evaluate the performance of data insertion and querying for both relational and graph databases. We observe that the relational databases show very good data insertion performance; however, the response time of backtrack queries is not acceptable as it takes over 20 minutes to extract causally dependent system objects for only a couple of files from a two-day system event log. On the other hand, graph databases show good response times for the same backtrack query as they store such relations explicitly in their graph format. On the other hand, their data insertion performance is often unacceptable to allow processing of real-time system events.

In this paper, we present \sysname, a graphical system for efficient loading, storing, processing, querying, and displaying system events to support computer forensics. \sysname effectively facilitates forensic analysis in a large enterprise.

\sysname flexibly offers the options of compressed in-memory storage, a traditional relational database system, and a graph database for storing the parsed audit logs. Because of their design nature, relational databases are highly optimized for storing tabular data; they may not perform well with a large number of chained joins. It can slow down operations that require an event stored in the database to be backtracked to its origin.
On the other hand, graph databases are designed with relationships in mind, making them much more efficient for backtracking relationships between events in a security log. However, graph databases tend to have low insertion performances, due to which they are unable to keep up with high-speed streams of logged data. Consequently, graph databases and in-memory storage are well suited for (near) real-time forensics when mini-batches of data are inserted, and fast query execution on graphs is needed. For post-mortem analyses, the relational database is more appropriate as it allows fast loading, indexing, and subsequent querying of large amounts of system events.

\sysname allows stored audit logs to be queried using a simple query language whose syntax and semantics are close to those of SQL. Importantly, and unlike SQL, this query language supports path queries and backtracking to an arbitrary depth from an identified resource. We note that the latter functionality is particularly crucial for successful forensics of prevalent computer attacks such as data exfiltration and kernel injections. User queries are parsed and interpreted by \sysname's query and visualization layer; then, the satisfying data is displayed as a color-coded graph or tree that can be rearranged, focused, and magnified for study.

\sysname belongs to a growing family of systems that support forensic analysis, which include Elastic~\cite{elastic}, AIQL~\cite{gao2018aiql}, SAQL~\cite{gao2018saql}, and Loglens~\cite{loglens}. However, these previous systems support simple keyword- and regular-expression based log data filtering only, and do not offer the useful backward trace query functionality. Furthermore, systems such as AIQL and SAQL are proprietary and not available for public use, in contrast with \sysname , which is open-source and free \footnote{\url{https://github.com/omid-s/cyber_deception/}} %

We evaluate \sysname on an extensive system call audit log of a real-world attack. We explore the performances of the various storage options as logs of increasing sizes are processed. Next, we evaluate the performance of executing various queries for forensic analysis, including those involving substantial backtracking. Our experiments reveal that \sysname scales linearly with the number of records processed. We point out queries for which one storage option is better than others. Finally, we briefly discuss a few case studies that illustrate the practical use of \sysname in computer forensics.

The rest of the paper is organized as follows. In section 2, we present a high-level overview of \sysname.
We describe details of three core layers of \sysname in the next three sections; we discuss the log ingestion layer in Section 3, the log storage layer in Section 4, and the query and visualization layer in Section 5. In section 6, we use large, real-world system logs to demonstrate the practical utility of \sysname. We summarize related work in Section 7, and Section 8 concludes the paper.

\section{Overview of \sysname}
\label{sec:sys_arch}
\begin{figure}
\centering
\includegraphics[width=0.8\columnwidth]{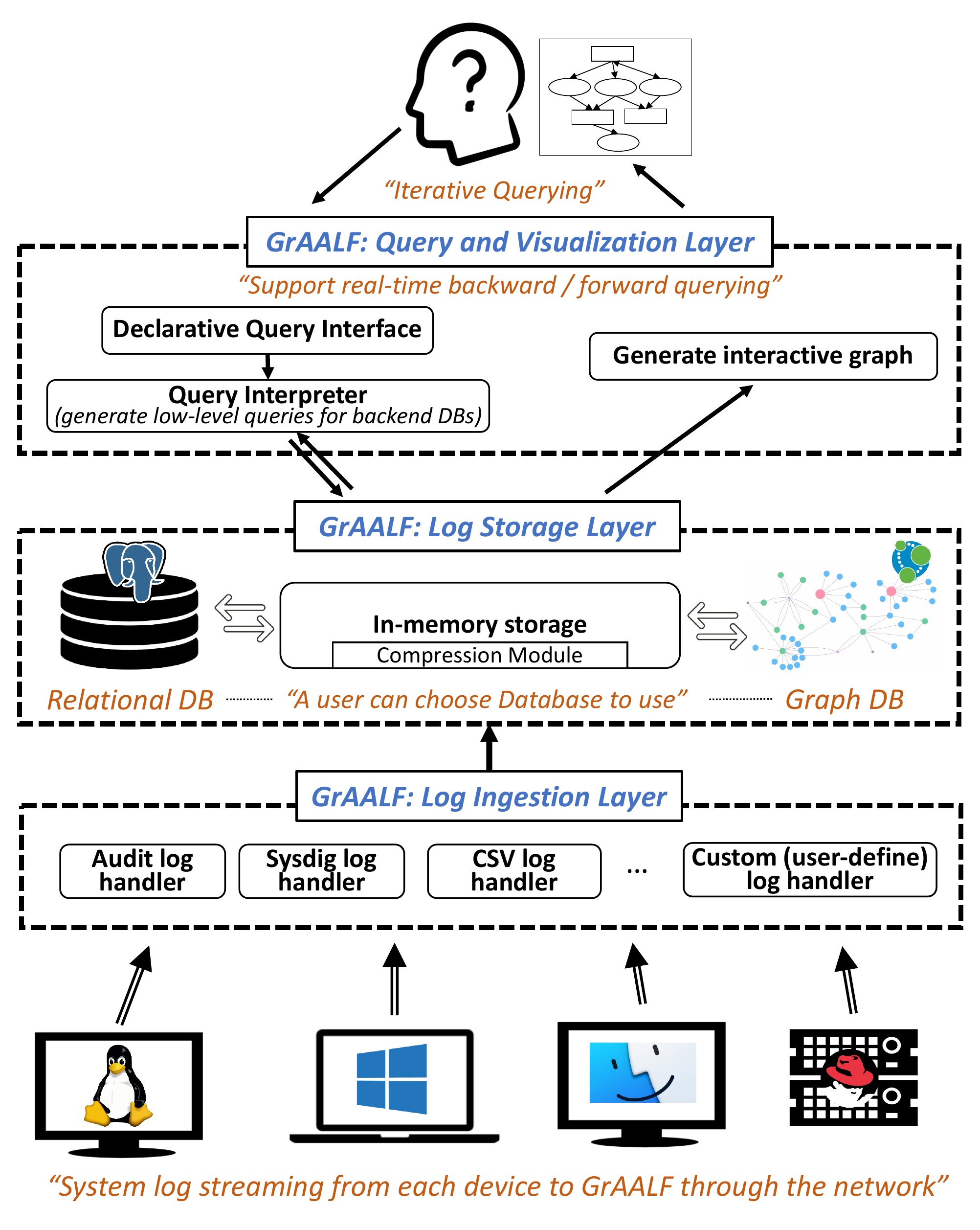}
\vspace{-1em}
\caption{A high level overview of \sysname.}

\label{fig:sys_arc}
\vspace{-1em}
\end{figure}

We illustrate the high-level overview of \sysname in Fig.~\ref{fig:sys_arc}. \sysname consists of three layers: log ingestion, log storage, and query and visualization layers. The log ingestion layer receives streaming system logs from hosts in the enterprise and processes them. The output of this layer will be formatted data that represent causal relations between system subjects (e.g., process, thread) and objects (e.g., file, network socket).

The log storage layer stores the output from the log ingestion layer into a permanent database. \sysname supports both relational and graph databases and allows a user to choose whichever is appropriate. \sysname also has in-memory buffer storage to enable the processing of enormous streaming data from multiple sources.

The query and visualization layer interacts with a user to receive queries and provide output as an interactive graph. The user can iteratively make queries based on the prior output graphs. The output graph visualizes causal relations in the system elements as well as interactions between different machines. More importantly, \sysname supports (near) real-time backward and forward queries through which the user can easily understand the origin of each system component and how one affects other components as the changes happen. We carefully design the storage layer and query interpreter for querying in-memory storage with the permanent database seamlessly. As soon as the ingestion layer processes the logs and delivers the output to in-memory storage, they are ready for querying.

The query and visualization layer also accepts queries to monitor the system for both stability and security purposes; this module will continuously monitor graphs produced by these queries and will notify the user when changes happen in any of them. This enables automated monitoring as well as automated detection of potential security incidents as they happen. 
\section{Log Ingestion Layer}
\label{sec:ingestion}

\begin{figure}
\centering
\includegraphics[width=0.8\columnwidth]{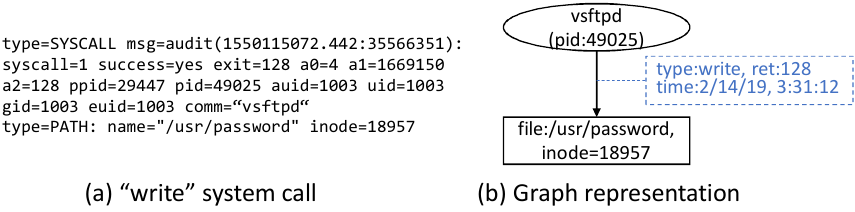}
\vspace{-1em}
\caption{System call event and graph representation.}
\label{fig:syscall}
\vspace{-0.5em}
\end{figure}

\begin{figure}
\centering
\includegraphics[width=0.8\columnwidth]{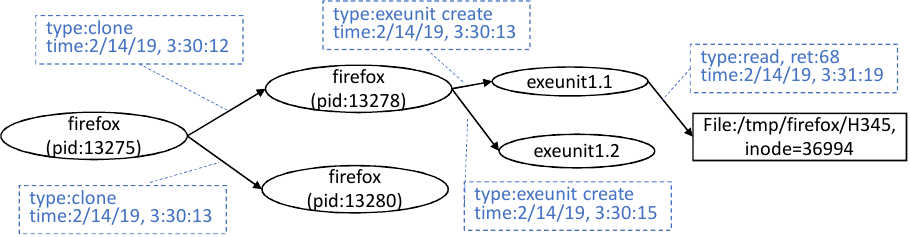}
\vspace{-1em}
\caption{Representing a new type of subject, execution unit.}
\label{fig:exeunit}
\vspace{-1em}
\end{figure}

System logs collected from hosts in the enterprise are streamed to \sysname's log ingestion layer. This layer contains several log handlers to support various types of system logs from different platforms.
For instance, most distributions of Linux and Unix systems use the Linux Audit tool~\cite{audit} to monitor system events such as system calls that reveal causal relations between system objects (e.g., process or thread) and subjects (e.g., file, network socket). On the other hand, Windows platforms mostly use Event Tracing for Windows (ETW)~\cite{etw} for system monitoring by intercepting system calls along with Windows API calls. DTrace~\cite{dtrace} is another popular tool that supports multiple platforms, including Linux, BSD, Solaris, and Windows. Commercial tools, such as Sysdig~\cite{sysdig} and Dynatrace~\cite{dynatrace}, are also available.
Also, various approaches~\cite{lee2013beep,pohly2012hifi,bates2015trustworthy,muniswamy2006provenance} have proposed to improve the effectiveness and efficiency of computer forensics.

Our main design goal in this layer is to provide extensibility and pluggability in the system. We aim to support any types of log events that can be represented as object nodes, subject nodes, and their relation edges.
For example, Fig.~\ref{fig:syscall} (a) shows a system call event in Linux where a process {\em ``vsftpd''} with pid 49025 reads a file {\em ``\textbackslash~etc\textbackslash~password''} with inode 18957.
This event is converted to a graph representation, as shown in Fig.~\ref{fig:syscall} (b). Nodes contain information about the process and the file, while the edges show how they are causally linked. In this manner, any system events that can be represented as graph nodes and causal relations between them can be ingested by \sysname.

Fig.~\ref{fig:exeunit} demonstrates how \sysname can support a new form of system objects and subjects.
A series of recent studies~\cite{lee2013beep, ma2017mpi, mci} demonstrates that default system logging approaches suffer from the \textit{dependency explosion} problem mainly due to long-running, event-handling processes. Process execution partitioning techniques~\cite{lee2013beep, ma2016protracer, ma2017mpi, mci, kcal} have been proposed to address the dependency explosion by introducing fine-grained system objects, called \textbf{execution units}, for both Windows and Linux platforms.
In this example, we adopt an execution unit as another type of system object. Additionally, we introduce an edge from a thread that includes the target execution unit to represent the relation between the two. This edge contains a timestamp of the unit's creation time.
Fig.~\ref{fig:exeunit} shows that the Firefox process with pid 13275 spawns two threads, 13278 and 13280, and 13278 creates two execution units and unit1.1 reads a file.
In this way, a user can easily add system objects and subjects by adding a definition of a new node, its fields, and edges from/to existing subjects and/or objects.

In this project, we implement the following three log handlers to process logs created by different systems:
\begin{itemize}[leftmargin=*]
    \item \textbf{Audit Log handler} processes logs generated by Linux Audit~\cite{audit}. It mainly contains system call events that contain process information (e.g., pid, parent process, user id, binary path, etc.), system object information (e.g., file name, inode, network socket address, child process information, etc.), and the description of event (e.g., system call number, arguments, return value, timestamp, etc).
    The format of the Linux Audit log is a key-value pair record that directly derives from system call arguments, so it often only provides a file descriptor instead of file name, path, inode, or network socket information. To address this, we maintain open file tables for each process to map file information, including an absolute path and inode, with a corresponding file descriptor. These tables are populated when we observe file open or other system calls that can manipulate the open file table (e.g., dup call to duplicate the open file descriptor). Then our handler converts an event into a graph format where subjects and objects become graph nodes, and an edge describes their relations and sends it to the storage layer.
    Additionally, we support modified Audit modules by ~\cite{lee2013beep, ma2017mpi} to include execution unit objects. We follow the same definition of execution units and unit dependency as the original author described in~\cite{lee2013beep}, and we maintain unit tables for each thread to identify the parent thread that created the execution unit and also maintain the memory dependencies with other execution units.
    \item \textbf{Sysdig handler} handles logs generated by Sysdig system monitor~\cite{sysdig}. Sysdig supports Linux, MacOS, and Windows. Similar to Linux Audit, Sysdig records system calls events. We support both output formats from Sysdig: json and plain text.
    Unlike Linux Audit, Sysdig provides target file and network information along with file descriptor number; that reduces our burden, so we can directly convert Sysdig logs into our graph format.
    \item \textbf{CSV handler} is a generic log processing module we provide where the user can define fields in a CSV (comma-separated value) header.
    We believe supporting CSV format is particularly useful because CSV is a well-known and popular format for storing and sharing data. It is easy to define the nodes, their fields, and relations between them in the CSV header, and it helps to map data in the CSV to the internal object structure.
\end{itemize}

If a user desires to use a custom logging system, 1) they can use CSV as an output format that \sysname can directly handle, or 2) they can implement a new handler that can parse events and pass the graph representation to \sysname.
Specifically, for each event, at least one object is required. Multiple objects and/or multiple subjects can exist in a single event. Each object and subject (i.e., node in a graph representation) has a unique signature (e.g., process id, file inode, etc. combined with the computer id and time) and unlimited fields for detailed information. These can be employed for user querying. An event can be represented as a directed edge between the object and the subject. There is no required field for edges, but we highly recommend providing a timestamp for each event to accurately identify indirect causal relations.
For example, assume that {\em process A} reads {\em file1} and then writes {\em file2}. After that, {\em process A} reads {\em file3}. We can infer that {\em file2} might be causally related to {\em file1} through {\em process A}. However, read events of {\em file3} happen after the write event, and thus there are no causal relations between {\em file2} and {\em file3}.
If we do not have timestamps on edge, we conservatively conclude that both {\em file1} and {\em file3} (potentially) affect {\em file2}, and this causes a false dependence.

\section{Log Storage Layer}

\sysname's log storage layer consists of two components: temporary in-memory storage and permanent database.
We consider two popular back-end storage systems: \textit{relational} database and \textit{graph} database systems. For the relational database, we mainly use \textit{PostgreSQL} in this paper but \sysname also fully supports \textit{mysql} and it can easily be extended to support \textit{Microsoft SQL Server} and \textit{Oracle} with minimal query adjustments.
While most traditional database implementations support {\it de facto} SQL language, newly emerging graph databases\cite{cypherisgood,francis2018cypher} do not have a standard language.
We use \textit{Neo4j} as a backend graph database, and \sysname works well with Neo4j's Cypher\cite{cypher} query language. We believe \sysname can be extended to support other graph databases compatible with Cypher, but we do not guarantee that.
To support a new graph database, \sysname requires a certain amount of modification to adopt the new structures and query language.

In addition to offering flexibility to the user to choose the favorite database system as the backend storage, we show performance comparisons to help a user choose the right one for their environment.

In general, queries about causal relationships between system subjects and objects perform better in a graph database because it does not require frequent table joins in processing the query. Graph databases directly store the relationships as edges between nodes and thus perform well in computer forensics.
We measure querying performance with the following two scenarios:
\begin{itemize}[leftmargin=*]
\item A forensic query from the user that returns an empty result. This query causes the database to scan the whole dataset but does not track any causal relationships.
\item A forensic query to backtrack the lineage of all files in {\em $/$home$/$} directory. This query causes the database to iteratively track causal relations for each file in {\em $/$home$/$}.
The output graphs contain all system nodes that directly or indirectly affect each file.
\end{itemize}

\begin{figure*}
\centering
\begin{subfigure}{.45\textwidth}
  \centering
  \includegraphics[width=0.8\columnwidth]{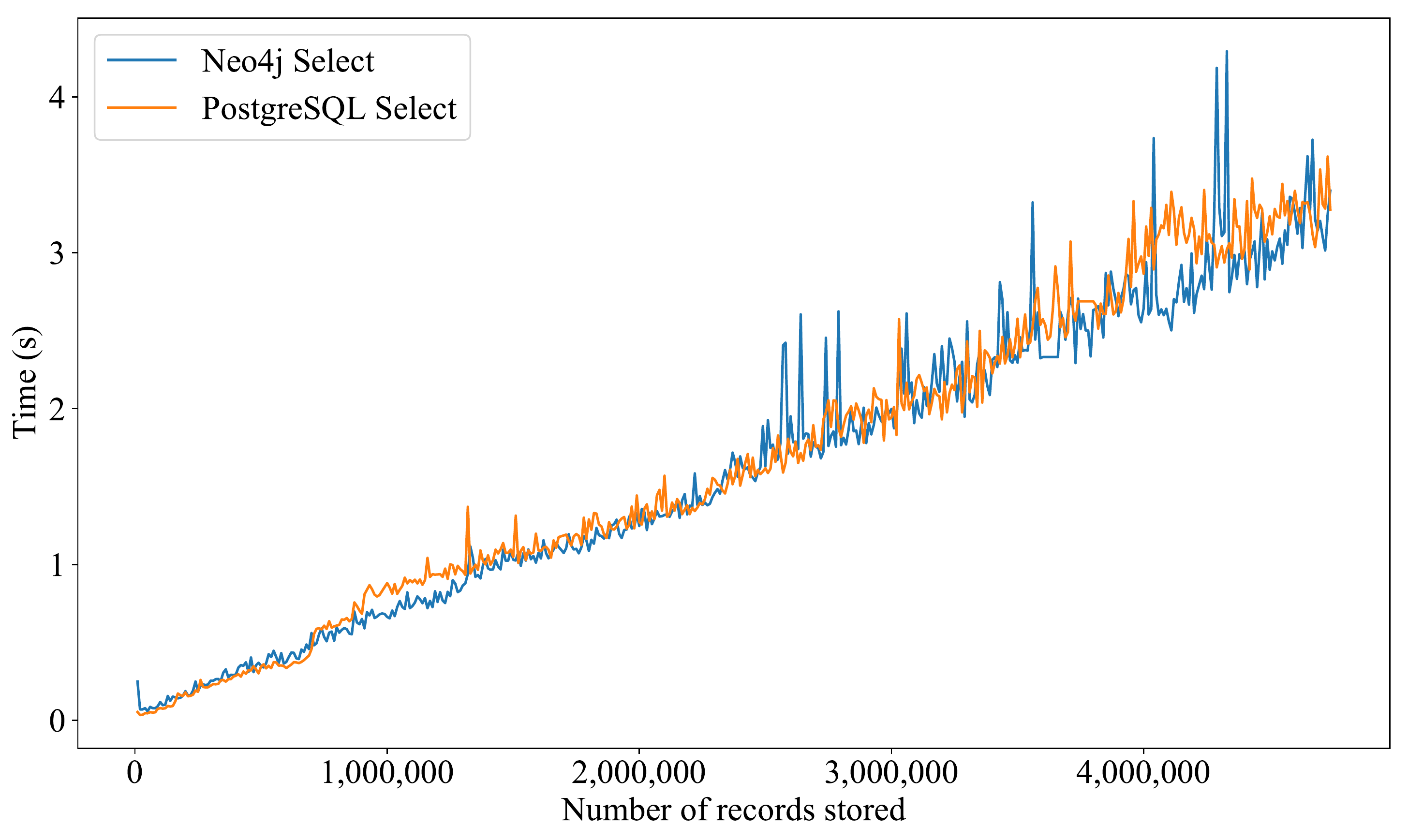}
   
  \caption{Query returns an empty result}
  \label{fig:empty_resutls_performance}
  \vspace{-1em}
\end{subfigure}%
\begin{subfigure}{.45\textwidth}
  \centering
  \includegraphics[width=0.8\columnwidth]{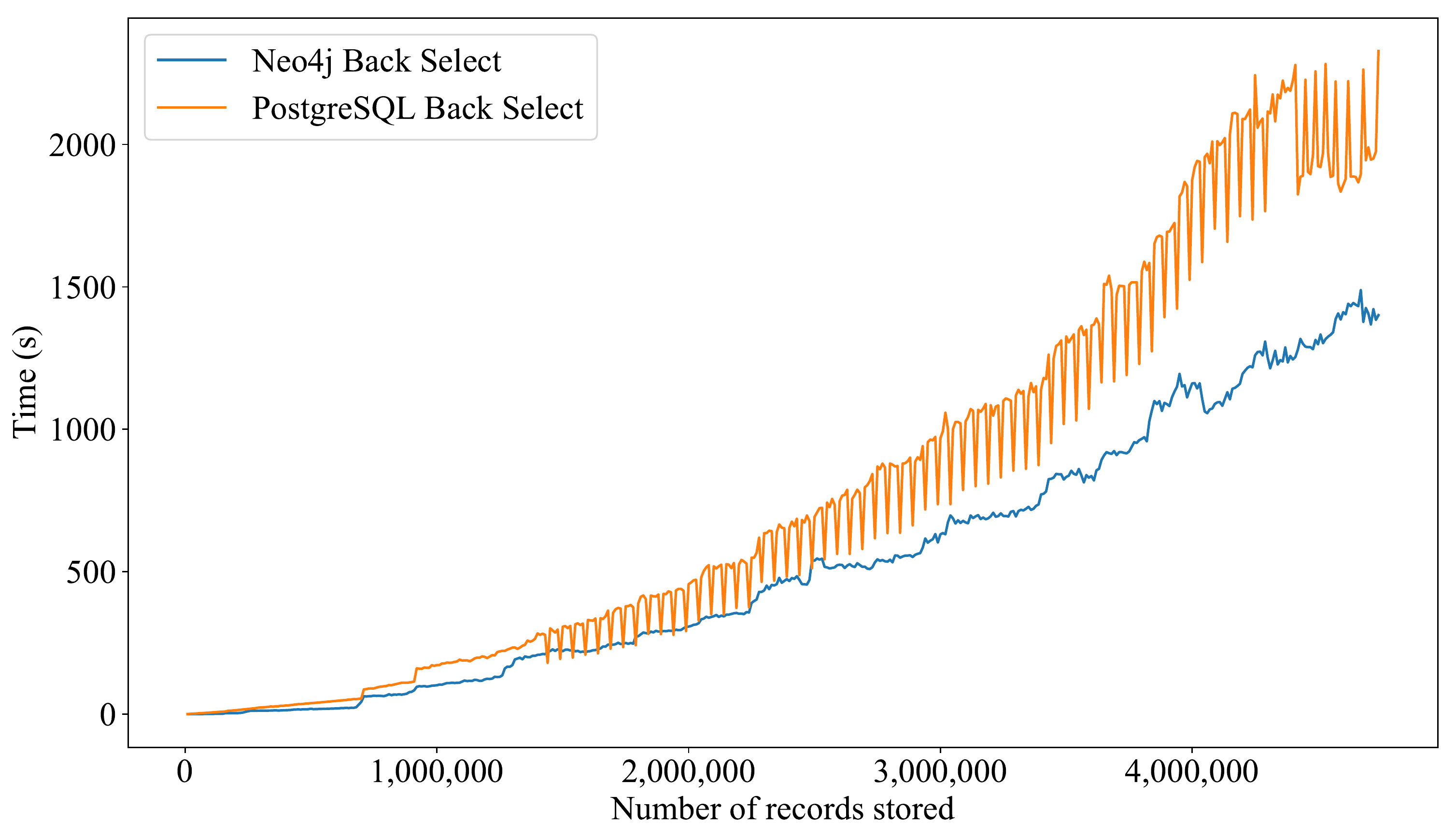}
   
  \caption{Backtrack the lineage of files in \textit{ /home/}}
  \label{fig:growing_results_performance}
  \vspace{-1em}
\end{subfigure}
\caption{Querying performance comparison between Neo4j and Postgres}
\label{fig:query_performance}
\end{figure*}

In this experiment, we use a real-world system log that we collect from a single host for 48 hours. The log is recorded by Linux Audit~\cite{audit} that runs on top of Ubuntu-16.04.
We execute each query with various sizes of datasets and measure the response time.
Fig.~\ref{fig:empty_resutls_performance} shows the performance when we execute the first query, which yields empty output. The performance of both databases is roughly linear to the size of the dataset. Note we intentionally use a query that returns an empty result to force the database to scan all items only once; thus, we can measure the performance of traversing the same set of data. We discuss the performance of dependency tracking in the next graph. In this evaluation, we do not observe any noticeable differences between PostgreSQL and Neo4j.
Fig.~\ref{fig:growing_results_performance} shows the performance of the backtrack query. It shows that the graph database (i.e., Neo4j) performs better for the query that requires tracking relationships.

\begin{figure*}
\centering
\begin{subfigure}{.45\textwidth}
  \centering
  \includegraphics[width=0.8\columnwidth]{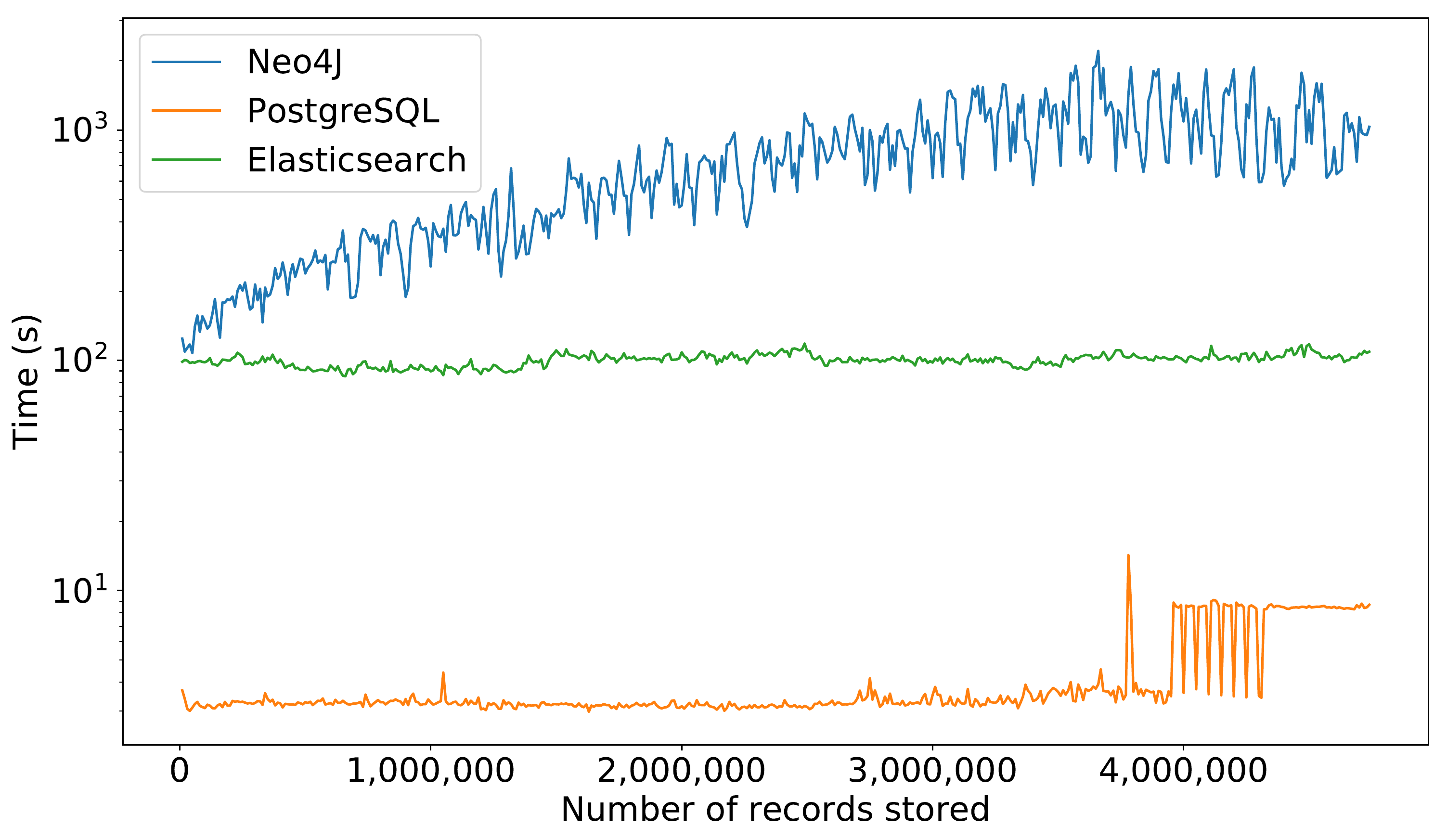}
   
  \caption{Insertion delay (Y-axis is in log-scale)}
  \label{fig:insertion_delay_performance}
  \vspace{-1em}
\end{subfigure}%
\begin{subfigure}{.45\textwidth}
  \centering
  \includegraphics[width=0.8\columnwidth]{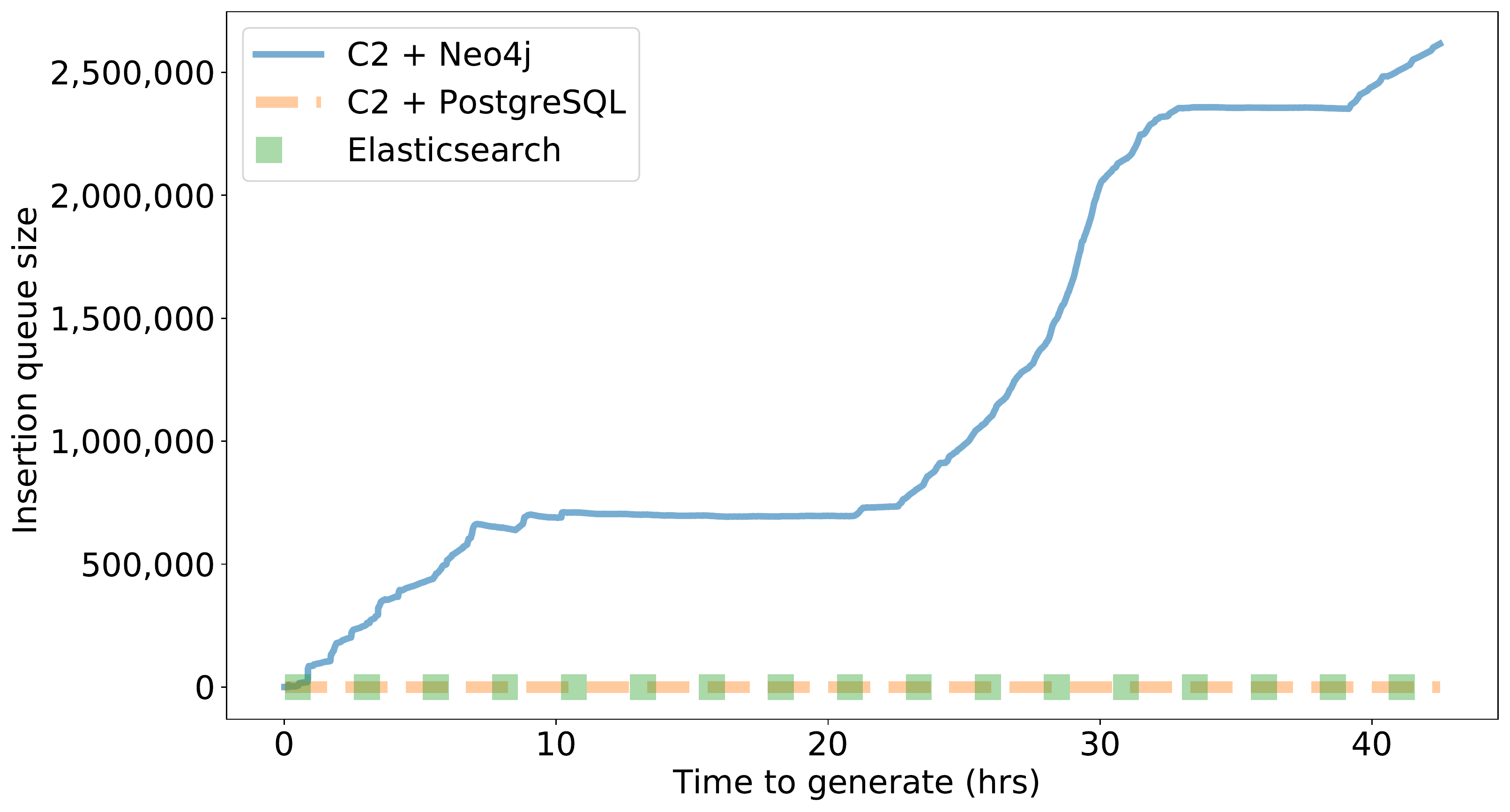}
 
  \caption{Length of insertion queue}
  \label{fig:insertion_queue}
  \vspace{-1em}
\end{subfigure}
\caption{Insertion performance comparison between Neo4j and Postgres and ElasticSearch}
\end{figure*}

On the other hand, considering data insertion performance, graph databases do not scale well with a large stream of data.
We insert system events in the log to each database according to the corresponding timestamp.
Fig.~\ref{fig:insertion_delay_performance} shows the insertion delay caused by each database. PostgreSQL can ingest all events without noticeable delay. However, Neo4j shows a severe delay. For instance, many events have to wait in the queue for more than 20 minutes. Fig.~\ref{fig:insertion_queue} shows the length of the insertion queue required to process all events for each database. After 32 hours of the experiments, Neo4j requires more than 2 million additional insertions to handle the logs from a single host, and it rapidly grows. It clearly shows that Neo4j does not provide enough insertion performance to handle real-world system events.

Note that in our evaluations, we do not try to optimize each database; instead, we use default configuration to show how they perform with the forensics queries as well as insertion of system event data.
For instance, both databases provide tools that can drastically speed up the insertions by potentially sacrificing the reliability of data. Neo4j offers a tool called \textit{batch insert}, but it bypasses multiple layers of validations and thus could lead to inconsistencies.
Additionally, there exist studies~\cite{gao2018aiql,gao2018saql,holzschuher2013graphDBperformance} that suggest methods for indexing and optimizing the data in different databases. Recently proposed high-performance databases~\cite{maass2017mosaic} can be used as well, but that is out of our scope in this project.

\subsection{In-memory Storage}
To address the insertion delay observed in Neo4j, and enable real-time querying for streaming inputs, we design and implement an in-memory storage system that can 1) maintain pending insertions in the system, 2) provide multiple graph compression techniques, 3) support seamlessly integrated querying between in-memory and database storage.

First, \sysname's in-memory storage acts as an event buffer where events can be queued for insertion into the backend database. Note that the user can query all events stored in the in-memory storage. After an event is inserted into the backend, we keep it in the in-memory storage only if the system has enough memory for better querying performance. We discuss details of in-memory querying in the next section.

Next, we introduce three graph compression techniques to reduce storage cost and improve querying and insertion performance. All compressions are done while the events are stored in the in-memory storage.
We have four options, as follows:
\begin{itemize}[leftmargin=*]
  \item \textbf{No Compression (C0): } This method returns all edges along with all of their metadata. We call it no-compression or C0. For instance, assume the system call event that Execution Unit ($E_1.1$) reads File \textit{'/tmp/firefox/H345'} ($F_1$) 3 times. Three read events will be stored as separate edges.
  \item \textbf{Lossless Compression (C1): } This approach merges all edges if their object node, subject node, and the relation type are the same. We merge edges into a single edge but keep all fields, including timestamps, in order to preserve all of the information that each edge contains.
  With this approach, the output graph has fewer edges, but the quality of information is the same as in C0. Applying C1 to the previous example, we merge all 3 edges into one but keep 3 timestamps.
  \item \textbf{Keeping Forensic Accuracy (C2): } This compression level also merges edges if their object, subject, and relation type are the same. However, unlike C1, we only keep the details of the first and last edges. With C2 used in the previous example, we merge 3 edges into one and keep only the first and last timestamps.
  This approach will lose some information (e.g., number of events between $E_1.1$ and $F_1$, total bytes that $E_1.1$ reads from $F_1$, etc.) but we maintain the accuracy of backward and forward tracking output.
  \item \textbf{Lossy Compression (C3):} In this case, we keep the first edge and discard the following edges if they have the same object, subject, and relation type. This approach will lose most of the ordering in events but still keeps the high-level causal relationship between nodes. More importantly, there is a chance that C3 causes bogus causality (i.e., false positives). For example, assume that $E_1.1$ reads from $F_1$ and later on, $E_2.2$ writes to $F_1$. With C0, C1, and C2, we can clearly tell that $E_1.1$ read $F_1$ before $E_2.2$ writes $F_1$, and thus, $E_1.1$ is not affected by $E_2.2$'s write event.
  On the other hand, if we adopt C3, we do not have enough information about which event happens before the other, and we should conservatively assume that $E_1.1$ might be affected by $E_2.2$ via $F_1$. This is a false causality introduced by the lossy compression.
\end{itemize}

\begin{figure}
\centering
\includegraphics[width=0.85\columnwidth]{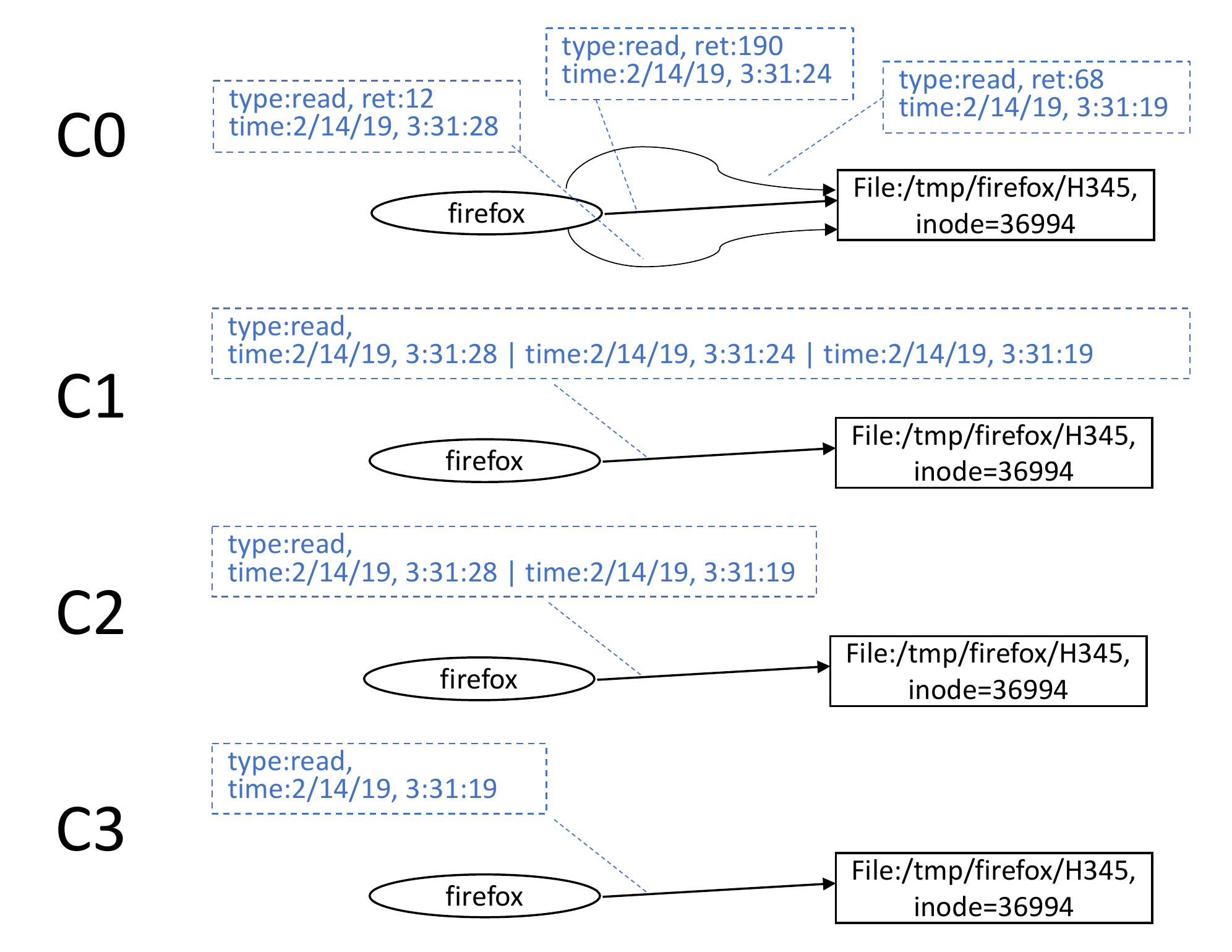}
\vspace{-1em}
\caption{Shows different compression modes - C0 keeps all edges with all their meta information, C1 will merge similar edges but keeps all timestamps, C2 keeps only the first and last timestamp, C3 keeps only the first timestamp.}
\label{fig:compression}
\vspace{-1em}
\end{figure}

We use all four compression levels to evaluate the insertion performance, and runtime, we allow the user to choose the compression level. We recommend C1 or C2 for most cases because C1 has the same quality of information as C0 with better performance, and C3 can cause false dependencies\cite{ma2015accurate}.
In addition to that, we provide two query modes: \textit{normal} and \textit{verbose}. In verbose mode, each event will be represented using a single unique edge so that C1's verbose output is the same as C0. The normal mode groups similar edges into one to make the resulting graph more readable. Furthermore, we plan to adopt recent log reduction techniques~\cite{lee2013loggc,ma2016protracer,xu2016highFidelity,hossain2018dependence} into \sysname as a part of graph creation or as an independent system to further reduce the size of the forensics graph. Fig.~\ref{fig:compression} shows the four compression models. 

\begin{figure}[!ht]
\centering
\includegraphics[width=0.8\columnwidth]{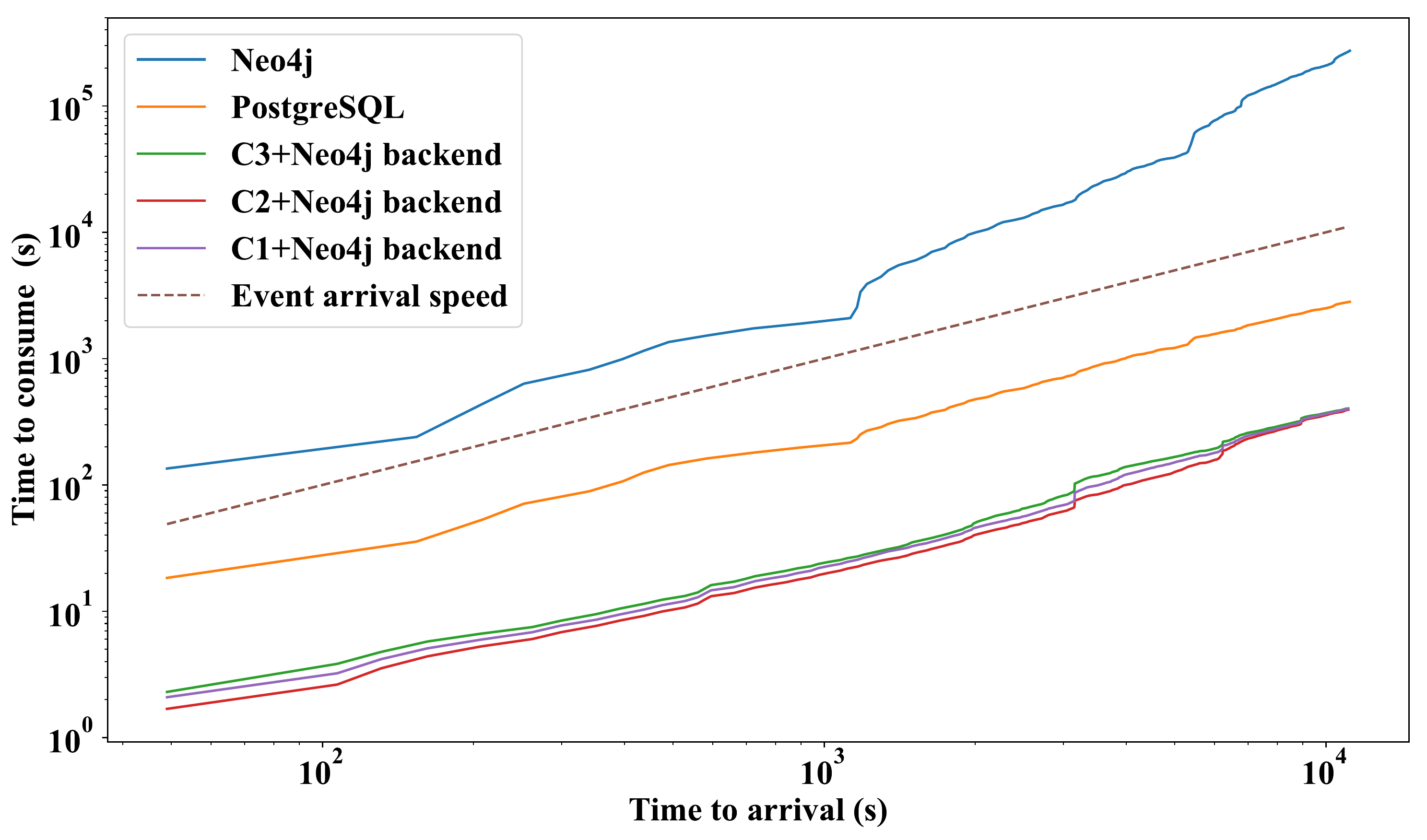}
\vspace{-1em}
\caption{Compares how different configurations can handle data in \sysname for a single machine stream (X-axis and Y-axis are in log-scale). }
\label{fig:processing_speed}
\vspace{-1em}
\end{figure}

Fig.~\ref{fig:processing_speed} shows insertion performance with different levels of compression. X-axis shows the event arrival time, and Y-axis presents the time when the event is consumed by \sysname. The dotted line shows the event arrival speed. If the insertion performance graph is above the dotted line, it means the processing is slower than the event arrival speed. If the performance graph lies below the dotted line, the processing speed is faster than event arrival and, thus, acceptable for real-time event processing.
As shown in Fig.~\ref{fig:processing_speed}, Neo4j alone is not acceptable for real-time event processing. Our in-memory storage with all compression levels can handle the data faster than event arrival speed, and the performance of C1, C2, and C3 do not show much difference.
In-memory storage with compression can improve PostgreSQL's insertion performance a little, but it is already faster than event arrival speed. We do not show PostgreSQL results for better readability.

\begin{figure}[!ht]
\centering
\includegraphics[width=0.8\columnwidth]{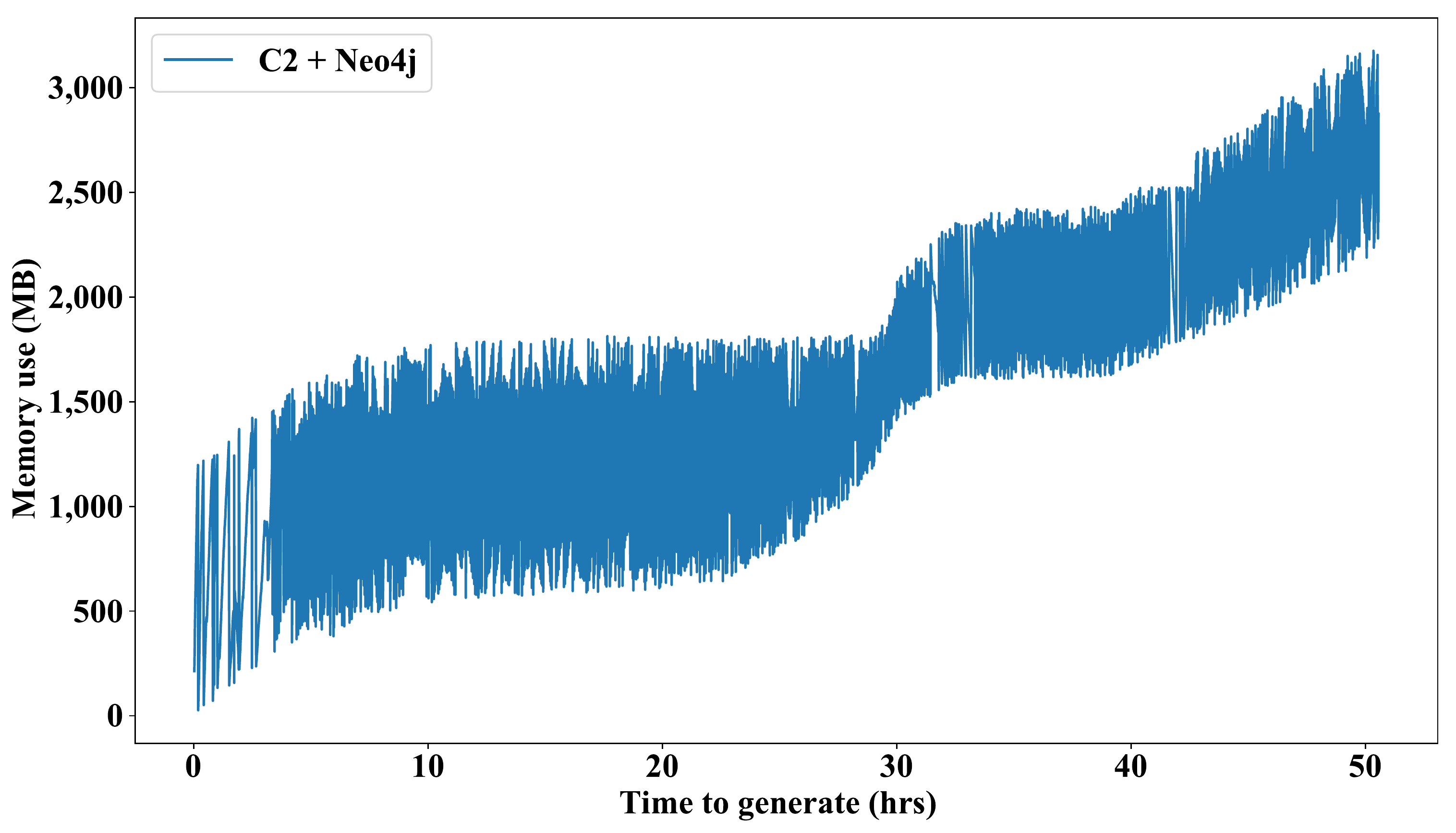}
\vspace{-1em}
\caption{Memory usage to store a forensics graph from a single host}
\vspace{-2em}
\label{fig:memory}
\end{figure}

Fig.~\ref{fig:memory} shows the amount of memory consumed by the in-memory storage process when we use C2 compression. It includes the space for graph buffer as well as the space for compression operations. The process occupies a large amount of memory for the compression, and frequently garbage collects to release the memory that is not required anymore, which explains the fluctuations in memory use.

\subsection{Graph Database Backend}

\sysname supports the storage of its graph in a graph database. To store records in a backend data store, and to minimize the memory used to keep them, we turn each record read from input into a small graph; the format of the log determines the size of this small graph. For example, if the graph is formatted as \textit{object affects subject} in each record; then, the subject and object will be considered as nodes, and "effect" will be represented as an edge. 
The created graph structure is passed to the graph database using a query that will ensure the nodes do not already exist in the graph and then will insert them; any preexisting node will be kept to preserve space and queryability.
Edge insertion, on the other hand, depends on the compression configurations. If C0 is selected, every edge is inserted in the database. However, with higher compression levels, an edge is inserted only once, and future occurrences of the same edge are considered updates that will make changes to the timestamps on the first edge. 
\sysname maintains a buffer of all the pending insertions. Buffer also checks all nodes and edges and prevents adding a node that already exists to the queue.

In order to improve insertion performances, multiple databases can be utilized, each with data for a particular time window.

\subsection{Relational Database Backend}

\sysname's relational data module supports PostgreSQL and MySQL database servers. Relational databases can be used in two forms, 1) the flat tabular format, 2) the two-table graph format.
In order to store data in a flat tabular format, we use one table with as many fields as in the internal record object. This opens room for direct storage and retrieval of those records; on the other hand, this comes at storage cost as some of the information will be copied multiple times. Also, as we will show in the evaluations section, querying information from a relational database for longer backtracking tasks is more time consuming compared to running the same query in a graph database.

The two-table graph model is implemented using two tables in a relational database: a vertex and an edge table. The nodes and edges have created the same as the graph database backend; however, nodes are inserted into the vertex table, and edges are inserted in the edge table separately. Queries sent to the backend for retrieving data in this format have both a higher number and more complexity than the graph database counterparts.

\vspace{-1em}
\section{Query and Visualization Layer}
\vspace{-0.5em}
\subsection{Query language}
\sysname's query language supports projection and filtering. The returned graph always starts from the nodes specified by the filter criteria and then applies the edge criteria if applicable.

The general structure of our query language is as follows :
\vspace{.5em}
\small
\fbox{\begin{minipage}{.98\columnwidth}
\textit{[+] [verbose] [[back]/[forward]] select {* ,[ Projection of Edge Types ]} from {*,[ projection of Node Types]} [where [[field] [operator] [value]] ] [;]}

\end{minipage}
}
\vspace{-1em}

\begin{itemize}[leftmargin=*]
\item \textbf{[Projection of edge types]:} this option is either * for all, or is a selection of event types for edges. For instance, if the dataset represents system call events, edge types can be one or more system calls (e.g., read, write, open, execve).

\item \textbf{[Projection of node types]:} this option is either * for all, or is a selection of node types (i.e., system objects and subjects). For instance, {\it file, process, socket, pipe}, can all be types of nodes for the system call datasets.

\item \textbf{[[field] [operator] [value]]:} these parameters can be any of the types described in the projection of node section. This part is optional, as you can skip it or use `any' to consider all types in the query processing. [field] can be any field of a node. For example, `pid' for a process id, `name' for a process name, and `uid' for a user id who launches the process can be used to selectively track process nodes in system call events.
[operator] can be  `is', `>' or `<' for exact matches or the `has' operator; arguments to the ``has'' keyword can also be POSIX regular expressions. Different criteria including ``and''; ``or" logical operators are also supported.
\end{itemize}

For instance, to identify all sources that affect ``/etc/active\_users.txt'', the user can compose the following backtrack query:
\vspace{-1em}
\begin{figure}[H]

\small
\centering
\fbox{\begin{minipage}{.9\columnwidth}
\textit{back select * from * where filename is `/etc/active\_users.txt'}
\end{minipage}}
\vspace{-1em}

\end{figure}

\noindent
Similar to other query languages, users can also issue queries to get or set environment variables and configurations. These environmental variables include the maximum number of edges to explore, whether to force the system to merge a new output into the previous graph; and display details of query execution, including the execution status of the storage layer and back-end database.

In addition, output graphs can be exported to various formats in other tools for further analysis. We currently support textual description, JSON format, and DOT, and \sysname can be extended to new formats. The ``describe'' keyword can be used to create a textual description of the output graph. The output will be the list of events that can be sorted by time or any field. Sort can be selected by issuing an ``orderby=[field]'' keyword in the describe query. ``format='' keyword sets the output format to `text', `json' or `dot'.

Automatic query tracking can be done by adding `+' before the query. When this feature is used, \sysname will notify the user when the results of the provided query have changed. Query identifier can be used with a `-' keyword to stop tracking.

\subsection{Query Interpreter}
\sysname translates a user query into one or more queries and fetches the corresponding data from in-memory storage and backend databases in the log storage layer.
This module runs as a separate thread to allow the user to interact with \sysname's console while the previous query is being processed.
As different backend databases use different query languages, we support both SQL-based and Neo4j's Cypher query languages. Regardless of the specific language used, we implement highly optimized in-memory query processing to allow the user to investigate real-time events.

\subsubsection{In-memory Storage}
All incoming queries are analyzed here, and outputs are quickly generated if the query can be fully answered without accessing backend databases.  
All outputs from backend databases will be transferred into in-memory storage and seamlessly correlated into a single graph. This allows \sysname to reduce graph construction efforts in other modules, while also providing a single interface to other modules to be able to consume the data.

The in-memory graph module maintains three hash maps to maintain indexes on nodes and edges. One map contains information in a nested map model, which starts with node type and then is mapped by their identifier. Another map uses a combination of identifiers from vertices to identify edges between every two nodes. The last hash map maintains a title-based map of the vertices to resolve regular expression and title matching queries faster.
Select queries will start based on their criteria; e.g., if the user is looking for processes, the search navigates to the type hash map, finds processes, and then looks for title matches. If no title is expressed in the query, all nodes in that type will be considered. To lookup nodes based on their titles; however, the search starts from the title hash map by finding keys that match the criteria and then processing their values.
Edge selection happens after nodes are selected. To find edges, we pick all combinations of two nodes and their corresponding edges from the edge map. If this search yields any results, those edges will be evaluated against edge criteria in the input query; e.g., if the user is asking for write system calls, returned edges should have `write' set as their system call. This helps efficiently find the desired nodes while maintaining a small overhead of nodes.

Back select queries start with a select query to find leaf nodes;
Then a Breadth-First Search traversal is done on the forensics graph, backtracking from leaf nodes. After the first layer, all nodes will have only one parent; while many processes might access one file or socket, a process cannot be created by more than one process. This assumption makes backtracking faster.
Forward select queries are designed to see the effects of node (system resource) on others. They start with finding the initial nodes in question. Once the initial nodes (root or roots) are found, BFS will be performed on the nodes.
Breadth-First Search is chosen as a uniform search strategy in graph traversals for two main reasons: 1) each record in a log input will contain information about multiple levels of the graph. Thus DFS has a higher chance of performing redundant queries. 2) especially in backtracking, the parents will be a line graph starting from the first of the second layer after the initial nodes. While this could benefit either DFS or BFS, we stay with BFS throughout \sysname to keep the model consistent.
To maintain the final query results, we use JUNG \cite{o2005jung} which provides a lightweight wrapper and user interface presentation for graphs and networks.

\subsubsection{Relational Database}
Similar to in-memory storage, once the input query is parsed, the SQL module will look for the nodes in the database. If process information is given as a criterion, then process, parent process, and ancestor process fields are considered in the initial query. When criteria are provided for a socket or file, file descriptor fields are the main source of filtering in the query. By default, \sysname stops running user queries with no criteria; 
Select queries will stop at this stage; all nodes that have been discovered will be sent to the memory module to create the final filtered graph.
Once seed nodes are discovered for a back or forward select query, BFS is performed on the database. As mentioned in the previous section, each record contains information about the file or socket as well as its process, parent process, and ancestor process; thus, each fetched record will create between one and four layers of depth in the final graph. For each new level in the graph, a new query is created and executed until there are no layers to add. While this requires many queries, \sysname tries to lessen this workload by merging queries for different paths of each layer into one, minimizing the round trip times and transaction overheads.

\subsubsection{Graph Database}

This is the simplest query modules. A graph database will already have the data stored in a graph model; thus, there is no need for tediously repeated queries. The graph-based module acts as an intermediary and translates \sysname query to Cypher queries. Cypher queries are sent to the Neo4j database, and the returned records are formed into a graph which will be appended to the existing graph in the memory. Most \sysname queries are translated into one optimized Cypher statement.

\vspace{-1em}
\section{Case Studies}
\vspace{-0.5em}

\subsection{Observation of Forensic Graph}

In this experiment, a graduate student used a Linux desktop for typical daily work for 48 hours. The user heavily used the Firefox web browser for surfing various websites. The user also frequently used text editors, a compiler, and productivity software working with documents and worksheets.

\begin{figure}[!ht]
\centering
\includegraphics[width=0.8\columnwidth]{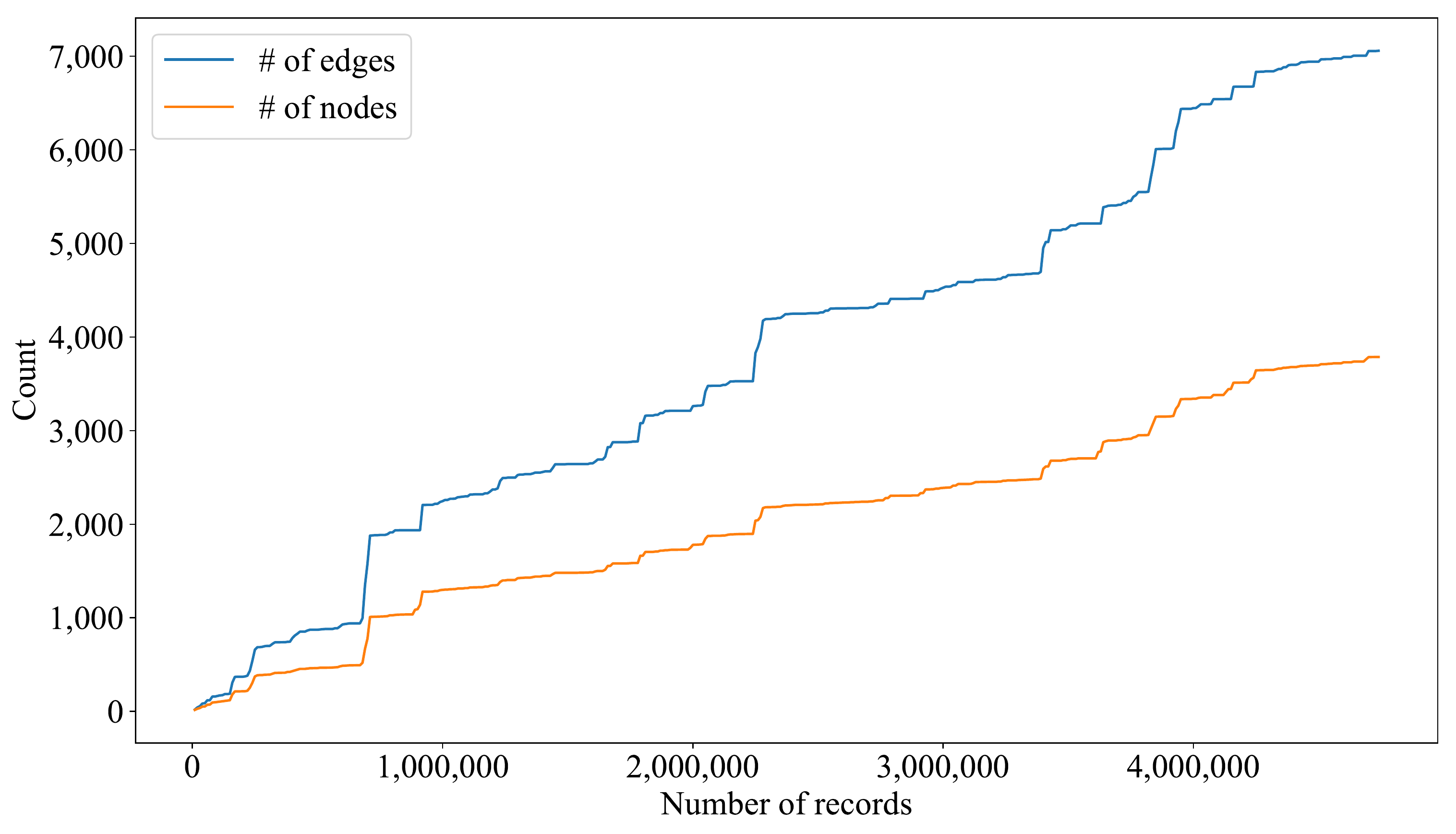}
\vspace{-1em}
\caption{The number of edges and nodes in a forensics graph for files in \textit{ /home/} directory on a Linux machine.}
\vspace{-1em}
\label{fig:back_select_node_change}
\end{figure}

To understand how forensic graphs evolve, we choose a subset of the data to study. We compose a query to identify files in the home directory (\textit{/home/}).
For each of these files, we count the number of related nodes and edges. We further execute backward tracking on all the nodes found in \textit{/home/} to find how their connectivity changes over time.
We count the number of nodes and edges for every 1,000 new event arrivals.
Fig.~\ref{fig:back_select_node_change} shows the trend.
Increasing connectivity is shown by the number of new edges introduced to the graph. The number of nodes increases as well, and that shows that the number of system subjects and objects related to the target files increases over time.
While the data we consider in this experiment is only a subset of the system logs, we choose the home directory as a considerable number of processes store files there and make changes to them.

\subsection{Case Studies}
In this section, we use four scenarios to demonstrate the practical usability of \sysname.
In this study, we use two realistic cyberattacks produced by the red team from the DARPA Transparent Computing program~\cite{darpatc}.
The system log contains over 5 billion records from multiple servers and desktop environments.
We demonstrate three cases where \sysname can help investigate cyber attacks and a real-time monitoring for early detection of malicious activities.

\begin{itemize}[leftmargin=*]
    \item \textbf{\textit{Attack Investigation 1}} : An administrator has notified that a suspiciously high volume communication has happened with a particular remote IP. The administrator starts the investigation with \sysname by backtracking the system events from the suspicious IP address. The final graph shows that the target machine has downloaded and run a compromised {\it Dropbear SSH} server, and it exfiltrates data from the victim machine.
    \item \textbf{\textit{Attack Investigation 2}} : An administrator is investigating a data exfiltration attack. An outside attacker has logged into the system via \textit{ssh} and transfers an important file to a remote host using \textit{csp} process.
    \item \textbf{\textit{Attack Investigation 3}} : This attack involves a backdoored FTP server application, \textit{vsftp}. The backdoor opens a particular port when an attacker logs into the FTP server with any user name, including a smiley face sequence of `:)'. The attacker exploits the backdoor functionality to drop a shell script that causes the server to restart.
    \item \textbf{\textit{System Monitoring}} : A system administrator is monitoring the enterprise system. The company does not allow execution of files directly downloaded from the Internet. In this scenario, the administrator uses \sysname to identify the violation of the policy effectively. 
\end{itemize}

\begin{figure}[H]
\centering
\includegraphics[width=.75\columnwidth]{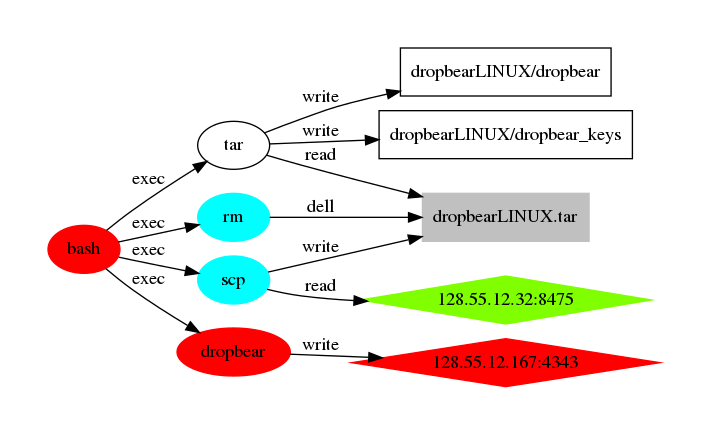}
\vspace{-1em}
\caption{Malicious drop and data exfiltration attack investigation graph. We use differently colored nodes to represent the steps of the investigation; \textit{red} nodes are added by the first query shown in Fig.~\ref{query:maldrop_and_exfil_queries}, \textit{white} nodes by the second, \textit{gray} ones by the third query, \textit{cyan} by the fourth, and \textit{green} nodes by the fifth query. It shows how a file has been dropped and installed on the victim machine, then used to exfiltrate data. }
\vspace{-1em}
\label{fig:maldrop_and_exfil_graph}
\end{figure}

\vspace{-1em}
\begin{figure}[H]
\small
\fbox{\begin{minipage}{\columnwidth}
Q1: \textit{back select * from soc where name has 128.55.12.167:4343} \newline
Q2: \textit{back select * from * where name is /dropbearLinux/dropbear;} \newline
Q3: \textit{forward select * from * where name is tar and pid is 13899;} \newline
Q4: \textit{back select * from * where name is dropbearLINUX.tar;} \newline
Q5: \textit{forward select * from * where name is scp and pid is 13870;}
\end{minipage}}
\vspace{-1em}
\caption{A set of queries used for data exfiltration attack.}
\vspace{-1em}
\label{query:maldrop_and_exfil_queries}
\end{figure}

\subsubsection{Malicious Drop and Data Exfiltration Attack}
The administrator has noticed a suspiciously large volume of traffic with the remote IP address 128.55.12.167 on port 4343 and starts to investigate it. In this attack investigation, 100GB of logs containing more than 500 million records are queried. %
Figure~\ref{fig:maldrop_and_exfil_graph} shows the detailed steps of how the investigation was performed, and the final attack graphs. 

\begin{itemize}[leftmargin=*]
    \item \textit{Step 1.}, The administrator, starts the investigation with the processes that have communicated with the reported IP/port endpoint.
    \item \textit{Step 2.} DropBear, a program alien to the environment and known for being a low profile tool, is seen to have sent information to the IP/port in question; this motivates the administrator to investigate the process. He then backtracks from the process to identify how and when it has created. From the query result, he observes a `tar' process has created the file.
    \item \textit{Step 3.} They further backtrack the file to `tar', then forward-tracks the file which has been extracted to create the `dropbear' executable. 
    \item \textit{Step 4.} They then find an `scp' process has downloaded the file, and then it was removed by an `rm' command.
    \item \textit{Step 5.} They, find the remote IP from which the program has been downloaded. At this point, the administrator can conclude, there has been a malicious download of a program that was used to send information to the suspicious IP. Since all the actions have initiated from a local shell, the administrator concludes it is an insider attack, or the attacker has physically accessed the victim machine.
\end{itemize}

In the investigation of this attack, the mean query response time is 16 milliseconds with a standard deviation of 13 milliseconds. AIQL~\cite{gao2018aiql} reports a minimum of 3.8 seconds for a similar task on a database of similar size.

\begin{figure}[!t]
\centering
\includegraphics[width=0.9\columnwidth]{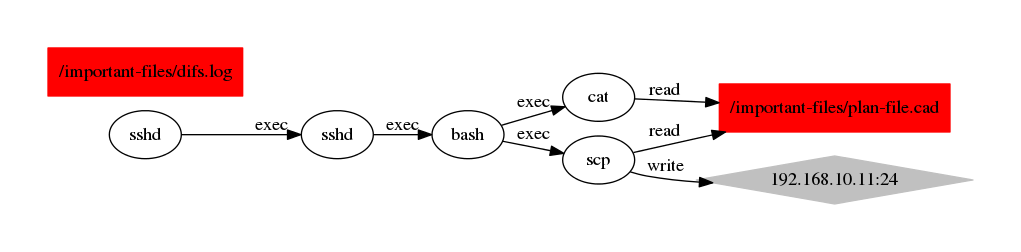}
\vspace{-1em}
\caption{Data exfiltration attack investigation graph;  \textit{red} nodes are added by the first query of Fig.~\ref{query:exfil_queries}, \textit{white} nodes by the second, and \textit{gray} ones by the third query. It traces the exfiltration of an important file  from the victim machine through a remote connection.}
\vspace{-1.5em}
\label{fig:exfil_graph}
\end{figure}

\vspace{-1em}

\begin{small}
\small
\begin{figure}[!ht]
\small
\fbox{\begin{minipage}{\columnwidth}
Q1: \textit{select * from file where name has /important-files/} \newline
Q2: \textit{back select * from * where name is /important-files/plan-file.cad;} \newline
Q3: \textit{forward select * from soc where name is scp and pid is 4667;}
\end{minipage}}
\vspace{-1em}
\caption{A set of queries used for data exfiltration attack.}
\vspace{-1em}
\label{query:exfil_queries}
\end{figure}
\end{small}

\vspace{-0.5em}
\subsubsection{Data Exfiltration Attack}
At the time of computer forensics work, the user has already been informed that an exfiltration attack has occurred. Thus, the goal of this investigation is to identify which files have been exfiltrated. 
Figure~\ref{fig:exfil_graph} shows the steps the user needs to follow in order to trace the attack.

\begin{itemize}[leftmargin=*]
    \item \textit{Step 1.} The user tries to find files that have been accessed during the given time period. Running the first query returns nodes in an important directory - for simplicity, we assume that the user has a list of the important files.
    \item \textit{Step 2.} Next, the user tries to discover what happened (i.e., who accesses, when it happens, how it has done) to the files by backtracking from them. Back select will provide the user with a hierarchy of processes that lead to accessing the file/files in question.
    \item \textit{Step 3.} The user has noticed a suspicious process (e.g., the user sees an \textit{scp} process in the back select from the last query). In order to determine where the file has been sent, the third query is issued. 

\end{itemize}

The final graph obtained via \sysname from the above 3 steps is shown in Fig.~\ref{fig:exfil_graph}.

\vspace{-1em}
\begin{figure}[!ht]
\small
\fbox{\begin{minipage}{0.9\columnwidth}
Q1: \textit{select * from * where name is myshell.sh} \newline
Q2: \textit{back select * from * where name is myshell.sh;} \newline
Q3: \textit{forward select * from * where pid is 24456 and name is sh;}
\end{minipage}}
\vspace{-1em}
\caption{A set of queries used for FTP exploit attack.}
\vspace{-1em}
\label{query:ftp_queries}
\end{figure}

\begin{figure}[!t]
\centering
\includegraphics[width=0.9\columnwidth,scale=0.9]{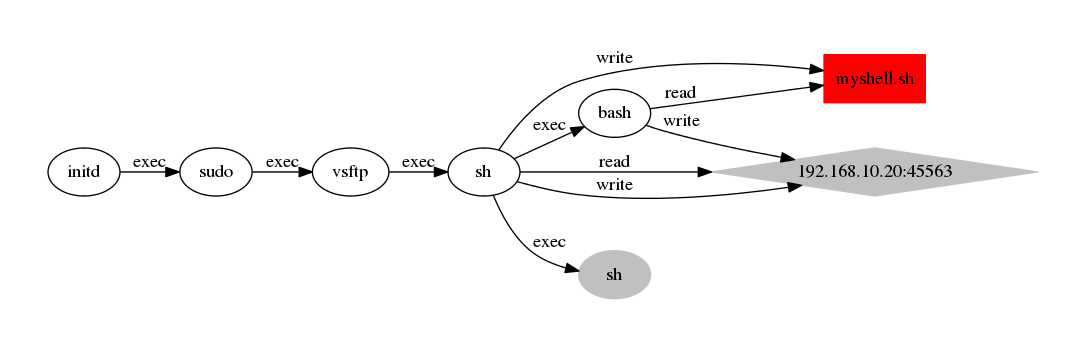}
\vspace{-1em}
\caption{FTP exploit attack investigation graph; \textit{red} nodes are added by the first query of Fig.~\ref{query:ftp_queries}, \textit{white} nodes by the second, and \textit{gray} ones by the third query. It shows how VSFTP's backdoor has been created and used to drop a shell file into the victim machine.}
\vspace{-1em}
\label{fig:ftp_graph}
\end{figure}

\subsubsection{FTP Backdoor Attack.}
The user has observed a sudden system reboot without any apparent reason. After the system turned on, the user detects that a new shell file is in the home directory; this information and the system logs are delivered to the forensics expert to investigate the incident. 
\begin{itemize}[leftmargin=*]
    \item \textit{Step 1.} The expert, locates the shell file.
    \item \textit{Step 2.} The expert, performs a back select from the file to see what processes led to the creation of the file. 
    \item \textit{Step 3.} The expert, finds a `sh' process which started a shell that wrote the shell file. Starting from the process, the expert performs a forward trace to find other malicious activities; this leads to a socket on a remote computer, with IP 192.168.10.20. 
\end{itemize}

Using the steps listed above, the expert can reconstruct the attack path. Based on the observed events, the expert concludes the intrusion was achieved by creating a backdoor from the "vsftp" server; accessing that backdoor, the attacker was able to drop a shell file into the system and execute it. 
Figure~\ref{fig:ftp_graph}  shows the steps to reconstruct the attack.

\subsubsection{System Monitoring.}
In this task, the administrator is monitoring the violation of the company's security policy. The company does not allow employees to download any programs directly from the Internet that can affect the other system objects (i.e., other processes or files). Fig.~\ref{fig:monitor_1_graph} shows results how such files are identified. 

\begin{itemize}[leftmargin=*]
    \item \textit{Step 1.} The user look for recent writes to the downloads directory.
    \item \textit{Step 2.} The user has the list of recently downloaded files. The next step is to find which one they choose to trace.
    \item \textit{Step 3.} The user chooses an executable file and performs a forward trace from that file.
\end{itemize}

\vspace{-1em}
\begin{figure}[!ht]
\small
\fbox{\begin{minipage}{0.9\columnwidth}
Q1: \textit{back select write from * where file name has /home /user1/Downloads/ and date has 2019-09-03} \newline
Q2: \textit{forward select * from * where name is /home/user1/Down-loads/dash} 
\end{minipage}}
\vspace{-1em}
\caption{Queries for the system monitoring task.}
\vspace{-1.5em}
\label{query:monitoring_queries}
\end{figure}

\begin{figure}[!ht]
\vspace{-1em}
\centering
\includegraphics[width=0.7\columnwidth]{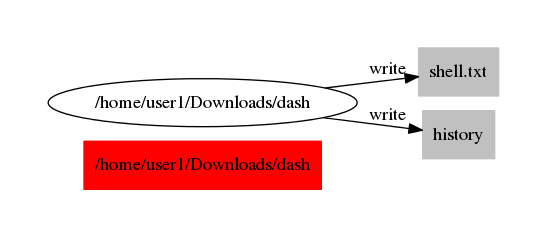}
\vspace{-1em}
\caption{Graph of the monitoring task. 
\textit{Red} nodes are added by the first query shown in Fig.~\ref{query:monitoring_queries},  \textit{white} nodes by the second, and \textit{gray} ones by the third query. It shows the monitoring of downloaded files.}
\vspace{-1em}
\label{fig:monitor_1_graph}
\end{figure}

With the steps mentioned above, the user will obtain a graph that shows the downloaded files and whether they have made changes to other files on that computer. Figure~\ref{fig:monitor_1_graph} visualizes the steps of this attack reconstruction in \sysname.

\section{Related Work}

Tools such as Elastic Stack are widely adopted in industry and academia for their agility and high performance when dealing with logs. Plaso~\cite{Plaso} and SOF ELK~\cite{sofelk} are two of the tools built on Elastic Stack; these tools have rich visualization and parsing features; however, they provide statistics-based reports, do not work with provenance graph models, and do not provide users with much queryabilty beyond filtering. These tools can be highly effective for monitoring or modeling the system state but are not as effective when it comes to provenance-tracking forensics querying. 

Other tools such as \cite{hassan2019nodoze,Plaso,CB} provide models based on artificial intelligence which will assist the forensics expert in monitoring the system and detecting malicious behaviors based on known patterns; however, these tools are not designed for manual forensics tasks such as whole system provenance tracking and are often bound to one scheme of proprietary data stream. This limits their application by a human analyst by 1) imposing limits, such as a limited number of system calls, and 2) limiting freedom of exploration while querying. 

\sysname is designed to give the user more flexibility in both data source and data exploration while maintaining features present in other works such as pattern matching on the streaming data and provenance tracking capabilities. Table~\ref{tbl:related_works} shows a comparison of \sysname with some related works by their capabilities. We compare these works by:
\begin{itemize}
    \item streaming analysis capabilities such as pattern matching or live monitoring and tracking denoted using attribute \textbf{S}.
    \item provenance backward and forward tracking denoted using \textbf{PG}; we use $\blacksquare$ when tracking on long sequences is possible and $\square$ when the tool provides tracking functionality, but it is not designed for long sequences from provenance graphs or can do tracking of events but does not provide a graph representation of provenance paths.  
    \item flexibility in the schema is the ability of the system to support more than one type of input denoted using \textbf{F}; e.g., Elastic Stack can process logs of arbitrary formats as long as they are provided in a key-value pair. 
    \item open-sourced and freely available denoted using \textbf{O}, where $\square$ represents tools with partially open-source or free versions. 
    \item support for a granularity smaller than the process denoted using \textbf{G}, which is unique to \sysname and helps prevent dependency explosion; other similar tools stop their analysis granularity at process or thread level, but \sysname supports finer-grained units.
    \item querying ability, denoted using \textbf{Q}, shows whether a tool provides an interface for users to sort through logs or not; $\blacksquare$ represents the presence of query languages, and $\square$ means there are some filtering capabilities but not a query language. 
    \item presence of automated anomaly detection capabilities denoted using attribute \textbf{A}. 
\end{itemize}

\begin{table}[!ht]
\centering
\vspace{-1em}
\caption{\sysname compared to related systems. $\times$ shows lack of support, $\square$ shows support with limited functionality, and $\blacksquare$  shows full support.}
\footnotesize
\begin{tabular}{c c c c c c c c }

\hline\hline %
    Title & S & PG & F & O & G & Q & A \\
    \hline
     \sysname     & $\blacksquare$ & $\blacksquare$ & $\blacksquare$ & $\blacksquare$ & $\blacksquare$    & $\blacksquare$ & $\times$ \\
    AIQL     & $\times$ & $\square$ & $\times$ & $\times$ & $\times$ & $\blacksquare$  &  $\times$   \\
    SAQL     & $\blacksquare$ & $\square$ & $\times$ & $\times$ & $\times$   & $\blacksquare$ &  $\times$  \\
    SOF ELK     & $\times$ & $\times$ & $\square$ & $\square$ & $\times$    & $\square$ &  $\blacksquare$ \\
    Carbon Black LiveOps     & $\blacksquare$ & $\square$ & $\times$ & $\times$ & $\times$ & $\square$    &  $\blacksquare$  \\
    NoDoze     & $\times$ & $\square$ & $\times$ & $\times$ & $\times$   & $\times$  &  $\blacksquare$ \\
    Plaso     & $\times$ & $\times$ & $\square$ & $\square$ & $\times$    & $\square$  &  $\blacksquare$ \\

    \hline
    
\end{tabular}

\vspace{-1em}
\label{tbl:related_works}

\end{table}

We discuss some of these related works in more detail below. 

SAQL \cite{gao2018saql} presents an online stream analysis tool that can detect predefined patterns in a fast stream of data. Patterns in this tool are defined using a query language that is capable of establishing time windows and calculating aggregations of events in those windows. The query language supports the definition of events separately, then connecting them to form a pattern. While the query language is very well designed to cover many tasks, the tool lacks constructs that focus on lineage of events (e.g., backward and forward tracking). Also, the query language is more complex than \sysname's.

AIQL \cite{gao2018aiql} presents a query tool that is capable of querying large sums of system provenance data. The query language is capable of backtracking from events in small steps through temporal orders. The system also provides a window-based aggregation model that allows users to query for abnormal behaviors, mostly in terms of frequency of events. The query language in this tool, much like in \cite{gao2018saql}, is not designed for tracing of long sequences of system events but instead focuses on events that are within a close circle of forensics steps. Also, the tool works by leveraging a highly optimized database backend.

Winnower \cite{hassan2018towards} is designed to cluster running tools like docker; it optimizes storage and reduces nodes by finding common graphs among different images. This relies on the fact that jobs running on these machines are heavily repetitive and similar. This allows the program to prune similar graphs from the network, which consists of many machines doing the same task. The tool also uses a query system capable of backward and forward tracking of provenance graphs with a Cypher-like query grammar. This tool is more oriented toward tasks with highly repeated components such as a computer cluster running the same or similar task; this idea does not apply to realistic program behaviors in an enterprise today.

Vast \cite{vallentin2016vast} focuses on a network-level log paradigm, but does handle logs in designing a distributed storage system that will keep and query network events. This system produces a complete graph of nodes in a distributed fashion where each node maintains an index and an archive of its own data. HDFS and bitmap index are being used to store the data and indexes. While \sysname can work with a distributed data store as well, the integrity of system logs prevents us from trusting a computer with its logs; thus, logs should be transported out as fast as possible. The query model in this work is also less flexible than that of the others.

Several causality analysis techniques exist~\cite{king2003backtrackingIntrusions, goel2005taserIntrusionRecovery, retro, king-ndss05,trailofbytes}, which use system call loggers to record important system events at run time and analyze recorded events to find causal relations between system subjects (e.g., process) and system objects (e.g., file or network socket).
For instance, BackTracker~\cite{king2003backtrackingIntrusions} and Taser~\cite{goel2005taserIntrusionRecovery} propose backward and forward analysis techniques in order to analyze system call logs and construct causal graphs for effective attack investigation.
Recently, a series of works~\cite{lee2013beep, ma2016protracer, ma2017mpi, mci, kcal} have proposed to provide accurate and fine-grained attack analysis. They divide long-running processes into multiple autonomous execution units and identify causal dependencies between them.
LDX~\cite{ldx} proposes a dual execution-based causality inference technique. When a user executes a process, LDX automatically starts a slave execution by mutating input sources. Then LDX identifies causal dependencies between the input source and outputs by comparing the outputs from the original execution and the slave execution.
MCI~\cite{mci} leverages LDX~\cite{ldx} to build a model-based causality inference technique for audit logs to infer causal relations between system calls.

\sysname can support all of the proposed logging techniques with the proper definition of subjects, objects, and relations. We can process their logs in real-time and provide \sysname's query interface to the user to enable interactive investigation and monitoring of the system.

\section{Conclusion}

We present \sysname, a graphical forensic analysis system for efficient loading, storing, processing, querying, and displaying of causal relations extracted from system events to support computer forensics.
\sysname offers the flexibility of choice between a relational database (e.g., PostgreSQL) and a graph database (e.g., Neo4j) for backend storage. \sysname's in-memory storage can be seamlessly integrated with either backend and provides (near) real-time forensic analysis of streaming system event logs.
\sysname{} provides a simple query language whose syntax and semantics are close to SQL. The user can compose a query to perform a backward and forward analysis to identify causal relations between system subjects and objects.
We use an extensive system call audit log that contains realistic attacks to demonstrate the practical utility of \sysname.
We will release the source code of \sysname as open-source software.

\section*{Acknowledgements}

This research is funded in part by grant \# W911NF-18-1-0288 from the Army Research Office. Authors would also like to thank Sean Frankum, Aditya Shinde and Muhammed AbuOdeh for their valuable inputs. 

\bibliographystyle{ACM-Reference-Format}
\bibliography{ref}

\end{document}